\documentclass[a4paper,nobibnotes,nofootinbib]{revtex4}

\usepackage{amsmath,amssymb}
\usepackage{epsfig}

\newcommand{\be}{\begin{equation}}
\newcommand{\ee}{\end{equation}}
\newcommand{\bL}{\begin{Large}}
\newcommand{\eL}{\end{Large}}
\newcommand{\ba}{\begin{eqnarray}}
\newcommand{\ea}{\end{eqnarray}}
\newcommand{\bc}{\begin{center}}
\newcommand{\ec}{\end{center}}
\newcommand{\bfig}{\begin{figure}}
\newcommand{\efig}{\end{figure}}

\newcommand{\g}{\gamma}

\newcommand{\rr}[4]{#1, {\it #2 \/}{\bf #3} #4}

\begin{document}
\title{Diffractive SUSY particle production at the LHC}

\author{M. Boonekamp}\email{boon@hep.saclay.cea.fr} 
\affiliation{Service de physique des particules, CEA/Saclay,
  91191 Gif-sur-Yvette cedex, France}
\author{J. Cammin}\email{cammin@fnal.gov} 
\affiliation{University of Rochester, NY, USA}
\author{S. Lavignac}\email{lavignac@spht.saclay.cea.fr}
\affiliation{Service de physique th{\'e}orique, CEA/Saclay,
  91191 Gif-sur-Yvette cedex, France\footnote{%
URA 2306, unit{\'e} de recherche associ{\'e}e au CNRS.}}
\author{R. Peschanski}\email{pesch@spht.saclay.cea.fr}
\affiliation{Service de physique th{\'e}orique, CEA/Saclay,
  91191 Gif-sur-Yvette cedex, France\footnote{%
URA 2306, unit{\'e} de recherche associ{\'e}e au CNRS.}}
\author{C. Royon}\email{royon@hep.saclay.cea.fr}
\affiliation{Service de physique des particules, CEA/Saclay,
  91191 Gif-sur-Yvette cedex, France, and Fermilab, Batavia, USA}

\begin{abstract}
  We give detailed predictions for the diffractive associated
  production of SUSY Higgs bosons and top squarks 
  at the LHC via exclusive double pomeron exchange mechanism. We
  study how the SUSY Higgs production cross-section and the
  signal-over-background ratio are enhanced as a function of $\tan
  \beta$ in different regimes. The prospects are particularly
  promising in the ``anti-decoupling'' regime, which we study in
  detail.  We also give prospects for a precise measurement of the
  top squark mass using the threshold scan of central diffractive
  associated top squark events at the LHC.
\end{abstract}

\maketitle

\section{Introduction}

The quest for supersymmetry (SUSY)  is one of the major goals in
high energy particle  physics, and is an important task for the  
experiments at the upcoming LHC. The discovery of SUSY Higgs scalar(s) and 
other new SUSY particles, among which the superpartners of the 
top quark are good candidates, would be an important achievement. Standard 
production mechanisms based on QCD are now well explored, at least for the main 
channels. However, due to both the general interest of the problem and some specific 
features of the SUSY Higgs and top squark sectors, one should seek for
alternative ways of SUSY production. 
In this paper we investigate the prospects for diffractive 
production 
of SUSY Higgs bosons and associated sparticles (stops)  in the central region 
of 
the detectors.

Standard Model Higgs boson production in double diffraction (denoted
DPE, for Double Pomeron Exchange) has already been studied in recent
years \cite{bialas1,bialas2,us1}, \cite{others,us,cox,sci,khoze}.
Many approaches have been pursued, considering diffractive scattering,
as in the Regge picture \cite{bialas1,bialas2,us1}, final state soft
colour interactions \cite{sci}, or fully perturbative exchange of
gluon pairs \cite{khoze}. We extend these studies to the SUSY Higgs
and sparticle sector in the framework of the Minimal Supersymmetric
Standard Model (MSSM)~\cite{MSSM}. This subject has already been
investigated in the literature
\cite{Cox:2003xp,Ellis:2005fp}\footnote{A preliminary study of the
  diffractive SUSY Higgs in our framework, emphasising the large
  enhancement factor at large $\tan \beta,$ can be found in
  Ref.\cite{Royon:2003ng}.}, and we will in particular focus on
central diffractive production of the lightest MSSM Higgs boson in the
``anti-decoupling'' regime which has not yet been studied in the
proposed framework. The ``decoupling'' and ``intense coupling''
regimes have been studied in Ref.~\cite{kaidalov03}, where it is shown
that diffractive Higgs boson production can help distinguishing
between $h$ and $H$ in the intense coupling regime.

\subsection{Central diffractive production of a heavy 
state}

One generally considers two types of DPE events for the production of a heavy 
state, namely ``exclusive'' DPE 
\cite{bialas1,bialas2,us1}, where the central heavy object is produced
alone, separated from the outgoing hadrons by rapidity gaps : 
\be p p \rightarrow p + \text{heavy object} + p\ , \label{exc}\ee 

\noindent and ``inclusive'' DPE \cite{others,us,cox,sci,khoze}, where the 
colliding Pomerons are resolved (very much like ordinary hadrons), accompanying
the central object with Pomeron ``remnants'' (X,Y):
\be p p \rightarrow p + X + \text{heavy object} + Y + p\ . \label{inc}\ee 

In general, exclusive production is considered most promising, because
of a good expected signal-over-background ratio (due to the large gaps
with no or low hadronic activity specific of diffractive events) and
because of the good missing mass resolution \cite{al00}~\footnote{The
  missing mass can be computed very precisely using roman pot
  detectors, and is equal to twice the mass of the heavy object in the
  case of exclusive events.}. Obviously, hard diffractive
cross-sections are of higher order than standard hard non-diffractive
ones, and this implies lower cross-sections.

There are \textbf{two objectives for this paper}: First, to present a detailed
calculation of the cross-section for exclusive DPE events, and second,
to elaborate on the advantage of using these events over other
processes to search for new heavy objects and to determine their
characteristics.
%% The purpose of 
%% this paper is to present a detailed evaluation of the cross-sections. 
Indeed, 
if the events are exclusive, \emph{i.e.}, if no other particles are produced 
in addition to the pair of heavy objects and the outgoing protons, the measurement
of the scattered protons in roman pot detectors 
allows to access the  Pomeron-Pomeron 
centre-of-mass \cite{al00}, and to study accurately the dynamics of the hard process. 
It is therefore possible to measure with great accuracy the properties
of new particles, e.g. their mass. It is also possible to
study the new
couplings by measuring cross-sections and angular 
distributions. As an example of this approach, we give a detailed description of the 
Higgs boson production cross-section and the top squark mass measurement at production threshold. 
The method can 
easily be extended to other heavy objects.

These studies rely on the \textbf{existence of exclusive events}.
At present, collider data from the Fermilab Tevatron give no evidence for
the existence of exclusive production processes. Only upper bounds are given \cite{cdfdijet}, 
dominated by the study of  dijet events near the kinematic limit of the 
produced diffractive mass. It is particularly difficult to identify purely exclusive
events experimentally, since the pomeron remnants are not detected, and the mass
resolution measured in the D\O\ or CDF main detectors suffer large uncertainties
due to the detector resolution. The evidence for exclusive events would be much
clearer in the $\gamma$ channel since the measurement of the diphoton mass fraction
does not suffer from these uncertainties. However, the production cross section of
those events is expected to be small at the Tevatron and LHC data will be necessary
to observe this process. Testing the present model with Tevatron data
will thus be challenging and 
LHC data in the beginning of data taking will be of fundamental importance.

\textbf{The theory of exclusive events} is uncertain as well. 
Different models lead to cross-sections at the LHC differing  by orders of 
magnitude. Two approaches are commonly discussed in the literature  
\cite{forshaw}, hereafter referred to as the  `Pomeron-induced' 
\cite{us1,us} and   
`proton-induced'
\cite{khoze} models\footnote{We  call `proton-induced' the model of 
Ref.\cite{khoze}, since it has the unintegrated gluon distribution inside the 
proton as the main ingredient. This is contrasted with the `Pomeron-induced' model 
\cite{us1,us} where the main source are  gluons  in the  Pomeron. The  
`proton-induced' and `Pomeron-induced' models are sometimes called respectively, 
the `Durham model' and the `Saclay model' for exclusive production in the 
literature \cite{forshaw}.} for exclusive 
production. 
They share   the common feature of   satisfying the upper 
bounds for exclusive production at the Tevatron and they give similar 
predictions 
for a low-mass standard Higgs boson cross-sections at the Tevatron. Apart from 
this, 
they come from different dynamics, since the `proton-induced' model is based on 
a  
semi-hard perturbative gluon mechanism, while the `Pomeron-induced' model is 
based 
on 
a soft, essentially non-perturbative, mechanism based on Pomeron exchange. 

\textbf{The `Pomeron-induced' model} is an extension \cite{us1} to the purely exclusive 
processes 
 of the original Bialas-Landshoff process for diffractive Higgs 
production \cite{bialas1} which was  applied to heavy quark pairs in 
\cite{bialas2}. In our model, both inclusive and exclusive diagrams come from
the same approach and are beased on the Bialas Landshoff model \cite{bialas1}.
The model starts with the 
same soft Pomeron exchange diagrams  (for ordinary dijet production  the $gg\to 
gg$ diagrams are also included \cite{bzdak})  and correct the 
result by  non-perturbative  rapidity-gap survival factors \cite{sp,Kupco:2004fw}. 
The 
energy 
dependence 
is related to the rise of ordinary hadronic cross-sections through
features of the soft Pomeron \cite{pom}.

An important issue of the   `Pomeron-induced' model is that the colourless Pomeron 
exchange implies the diffractive phenomenon.
Our model follows the Bialas Landshoff model presciption, namely we assume the
existence of a direct coupling between the Pomeron and the heavy mass object
which is produced (Higgs or jets for instance).
Hence, by definition it takes into account the veto 
on gluon radiation from the production mechanism. In other words, since we consider a
pomeron induced model, there is no further need for a Sudhakov form factor to suppress
radiation at the pomeron level.
The counterpart is that it involves 
non perturbative mechanisms which have to be modelled \cite{bialas1} using non 
perturbative gluon propagators. Hence, 
in the  `Pomeron-induced' framework,  the normalisation is not determined 
theoretically, since it is related to the unknown non-perturbative strong 
coupling 
constant $G^2/4\pi$ \cite{bialas1,bialas2}. However, the coupling constant 
value 
can 
be fixed phenomenologically by requiring consistency with the description of the 
inclusive 
dijet production within the same scheme, and is found \cite {us} 
compatible with the value $G^2/4\pi=1$ chosen in  \cite{bialas1,bialas2}. For 
our 
study, we will use this value to evaluate the production cross-sections

\textbf{The `proton-induced' model} is based on the perturbative calculation
related to the suppression of QCD radiation due to Sudakov
form-factors. Besides, soft radiation is forbidden through a rapidity
gap survival formalism \cite{sp}. The same Sudakov form-factors are
responsible for a damping of large mass diffraction (for Higgs boson
production, it is compensated by a rapid growth as a function of
energy). In particular, one expects to get a negligible top squark
pair exclusive cross-section \cite{durhamtop}. Indeed, in the
`proton-induced' model approach, the normalisation is estimated purely
theoretically (except the lower bound constraints on dijet production)
and leads to a top squark cross-section too small to be observed at
the LHC.

%% The problem of normalisation of cross-sections is naturally of
%% relevance for our study. In the `proton-induced' model, the
%% normalisation is theoretically estimated through the perturbative
%% calculation and would also lead to a top squark cross-section too small to
%% be observed at the LHC.

We will focus on exclusive production (\ref{exc}) and restrict ourselves to the
original Bialas-Landshoff type of models \cite{bialas1,bialas2,us},
with their extension to SUSY Higgs and stop production which we will
develop in the next sections. There are large theoretical
uncertainties of diffractive production which are related to the
non trivial interplay between perturbative and non perturbative
contributions. We found it instructive to use the non perturbative
`pomeron induced' model and compare it to the predictions for the
`proton-induced' model. 
%% The latter has strong suppression of high mass
%% production due to Sudakov form-factors \cite{kaidalov03}. 
Indeed,
it will be straightforward to extend these studies to other
models. A recently developed Monte-Carlo program, {\tt
  DPEMC} \cite{dpemc}, implements the models of
\cite{bialas1,bialas2,us1,cox,khoze}. Moreover, most of the plots are
in terms of s/b and enhancement factors, that are ~independent of the
models.

Even though inclusive DPE (\ref{inc}) is a less promising search 
channel, it is still important to consider. In particular, it 
is interesting to evaluate  the tail of the inclusive mass spectrum 
(``quasi-exclusive'' processes) since it 
constitutes a background to exclusive DPE. 
In addition, only inclusive DPE has actually been observed for high 
central masses~\cite{cdfdijet}. Issues of inclusive DPE will be
discussed in a future publication.

\subsection{The relevant SUSY spectrum: SUSY Higgs boson and top squarks}

Due to the limitation in the available total energy for production, 
diffractive production is favoured for the production of  SUSY particles 
in 
the lower range  of their mass in the  admissible set of model parameters. 
Hence, 
we will focus on this range for the SUSY Higgs and top squarks sector.
The regions of the MSSM parameter space 
that favour a light Higgs boson and those that favour a light top squarks are not the same, 
so 
a specific study is required separately for  
Higgs bosons and top squarks, see Section~III and IV.

It is well-known that the Higgs boson sector of the MSSM
is richer than that of the Standard Model.
First, it contains five physical scalar degrees of freedom, instead of a single 
one: two 
CP-even neutral Higgs bosons $h$ and $H$, a pseudoscalar Higgs bosons $A$ and a 
charged Higgs boson pair $H^\pm$. Secondly, the lightest MSSM Higgs boson $h$ 
may look 
very different from the SM Higgs boson. One can define (at least) 
three 
noteworthy regimes for the couplings of the neutral MSSM Higgs bosons $h$, $H$ 
and $A$:

(i) the {\it decoupling} regime, in which $h$ behaves like the SM Higgs
boson~\cite{haber93,gunion02};

(ii) the {\it intense coupling} regime, in which the couplings of all three 
neutral 
Higgs bosons are very different from those of the SM Higgs boson~\cite{boos02};

(iii) the so-called {\it anti-decoupling} regime \cite{djouadi05}, in which  
$H$ behaves like the SM Higgs boson, while $h$ has enhanced (resp. suppressed) 
couplings to down-type fermions (resp. up-type fermions and gauge 
bosons)~\cite{gunion96}.

It is also well-known that the MSSM Higgs boson sector contains at least one scalar 
$h$ 
with rather low mass (the other,  $H$, being with larger but possibly 
accessible 
mass) which  gives a particular interest for diffractive production as we will 
study in detail in this paper. Indeed, the small mass, the sometimes small 
rate
and the experimentally difficult standard channels, e.g., the decay
into $\g \g$,  
enhances the interest in alternative production modes and decays such as
diffractive production. We will in particular 
focus on central diffractive production of the lightest MSSM Higgs boson  in 
the 
anti-decoupling regime (iii) which has not yet been intensively studied in the 
proposed framework. The decoupling and intense coupling regimes have been
studied in Ref.~\cite{kaidalov03}, where it is shown that diffractive Higgs 
boson
production can help to distinguish between $h$ and $H$ in the intense
coupling regime.

The interest of MSSM Higgs boson production via exclusive diffractive 
production 
parallels a similar analysis for the Standard Model Higgs boson, with some 
distinctive 
features which enhance the specific  production and branching modes. 

\vspace{.5cm}

The discovery of ``sparticles'' at the LHC would be the clearest and most 
exciting signal of new fundamental physics beyond the Standard Model. Among 
these, the scalar superpartners of the top quark are expected to be those with 
smallest mass among scalars in a large portion of the MSSM parameter space.
Indeed, various supersymmetric scenarios can accommodate a light top squark 
consistent
with the experimental bounds on other sparticle masses and with measurements of
observables that could be affected by large supersymmetric contributions, such
as the anomalous magnetic moment of the muon or the branching ratio of the
flavour violating decay $b \rightarrow s \gamma$. Minimal supergravity
\cite{SUGRA} (mSUGRA) scenarios with a light top squark typically require a low 
gaugino mass
parameter $m_{1/2}$ and a large $A$-term parameter $A_0$. The need for a
small $m_{1/2}$ is due to the fact that the renormalisation
group equations for the soft supersymmetry breaking squark masses $M^2_Q$
and $M^2_R$ receive a large contribution from gluinos: the larger $m_{1/2}$,
the higher the weak-scale values of $M^2_{Q_3}$ and $M^2_{U_3}$. As an example,
the Snowmass Point 5 (SPS 5), defined by the following values of the mSUGRA
parameters: $m_0 = 150$~GeV, $m_{1/2} = 300$~GeV, $A_0 = - 1000$~GeV,
$\tan \beta = 5$ and $\mbox{sign} (\mu) = +$, yields the following top squark 
and 
sbottom
spectrum \cite{ghodbane02}, as generated by the program SUSYGEN 3.00/27
\cite{SUSYGEN}:
\be
  m_{\tilde t_1} = 210\ \mbox{GeV}\ , \qquad m_{\tilde t_2} = 632\ \mbox{GeV}\ 
, 
\qquad
  m_{\tilde b_1} = 561\ \mbox{GeV}\ , \qquad m_{\tilde b_2} = 654\ \mbox{GeV}\ 
.
\ee
For comparison, a ``typical'' mSUGRA scenario with vanishing $A_0$, the 
``post-WMAP
benchmark scenario'' B', defined by $m_0 = 60$~GeV, $m_{1/2} = 250$~GeV, $A_0 = 
0$,
$\tan \beta = 10$ and $\mbox{sign} (\mu) = +$, yields  \cite{battaglia04}:
\be
  m_{\tilde t_1} = 393\ \mbox{GeV}\ , \qquad m_{\tilde t_2} = 573\ \mbox{GeV}\ 
, 
\qquad
  m_{\tilde b_1} = 502\ \mbox{GeV}\ , \qquad m_{\tilde b_2} = 528\ \mbox{GeV}\ 
.
\ee
Light top squarks and bottom squarks \footnote{In the following, we consider 
only top squarks.
But the study remains unchanged for squarks (\emph{i.e.}, bottom squarks) if 
their masses are
sufficiently
low.} can also arise from non-minimal SUGRA models,
e.g. from scenarios with an inverted mass hierarchy in the squark sector 
\cite{inverted}.
Assuming $M^2_{\Phi_3} \ll M^2_{\Phi_{1,2}}$ ($\Phi = Q, U, D$) at the GUT 
scale
and small gaugino masses, one ends up with very low third generation squark 
masses
at the weak scale due to strong 2-loop renormalisation group effects 
proportional to
$\alpha^2_S \mbox{Tr} (2 M^2_Q + M^2_U + M^2_D)$ \cite{arkani97}. On
the contrary, the first two 
squark
generations remain heavy. The top squark and
sbottom squared masses can even be driven negative if the
GUT-scale hierarchy $M^2_{\Phi_3} \ll M^2_{\Phi_{1,2}}$ is too
pronounced, or if the gluino mass, whose contribution to the running of the 
squark
masses tends to compensate for the two-loop gauge contribution, is too small.
  
The paper is organised in the following way. In the next Section
II, we introduce the concept of central diffractive production of SUSY
Higgs bosons and top squarks; in II-A, the formalism of exclusive production
and in II-B the experimental context are presented. In section III, we
focus on the MSSM Higgs boson sector; in III-A, theoretical aspects of the
Higgs boson spectrum and in III-B, the predictions for the LHC are
displayed. In section IV, the case for top squark, and eventually bottom squark
production is discussed; in IV-A, the theoretical framework, in IV-B,
the predicted cross-sections and missing mass distribution and in
IV-C, the top squark mass measurement by a threshold scan are given. The
paper ends by a conclusion and outlook.

\section{Diffractive production of SUSY Higgs boson and top squarks}

\subsection{Exclusive central diffractive production}

In this section we introduce the model~\cite{bialas1,bialas2,us} that
is being used to describe 
exclusive MSSM Higgs bosons and top squark pair production in double 
diffractive 
production. 
In \cite{bialas1,bialas2}, the diffractive mechanism is based on two-gluon 
exchange 
between 
the 
two incoming protons. The soft pomeron is seen as a  pair of 
gluons 
coupled non-perturbatively
 to the proton. One of the gluons is then 
coupled 
perturbatively to the hard process, either the SUSY Higgs bosons, or 
the $\tilde t \bar{\tilde t}$ 
pair, while the other one plays the r\^ole of a 
soft 
screening of 
colour, 
allowing for diffraction to occur.
The corresponding cross-sections for Higgs bosons and $\tilde t \bar{\tilde t}$ 
production 
read:

\begin{eqnarray}
 d\sigma_{h}^{exc}(s) &=& C_{h}\left(\frac{s}{M_{h}^{2}}\right)^{2\epsilon} 
\delta\left(\xi_{1}\xi_{2}-\frac{M_{h}^{2}}{s}\right)
\prod_{i=1,2} \left\{ d^{2}v_{i} \frac{d\xi_{i}}{1-\xi_{i}} \right.
\left. \xi_{i}^{2\alpha'v_{i}^{2}} \exp(-2\lambda_{h} v_{i}^{2})\right\} 
\sigma (g g \rightarrow h)
 \nonumber \\
d\sigma_{\tilde t \tilde{\bar{t}}}^{exc}(s) &=& C_{\tilde t \tilde {\bar{t}}} 
\left(\frac{s}{M_{\tilde t \tilde{\bar{t}}^{2}}}\right)^{2\epsilon}
\delta^{(2)}\left( \sum_{i=1,2} (v_{i} + k_{i}) \right)
\prod_{i=1,2} \left\{ d^{2}v_{i} d^{2}k_{i} d\xi_{i} \right.
d\eta_{i}\  \xi_{i}^{2\alpha'v_{i}^{2}}\! 
\left. \exp(-2\lambda_{\tilde t\tilde{\bar{t}}} v_{i}^{2})\right\} {\sigma} 
(gg\to {\tilde t \tilde{\bar{t}}}\ )
\label{exclusif}
\end{eqnarray}
where, in both equations,  the variables $v_{i}$ and $\xi_{i}$  
denote the transverse
momenta and fractional momentum losses of the outgoing protons. In the second 
equation, 
$k_{i}$ and $\eta_{i}$ are the squark  transverse 
momenta and  rapidities. $\sigma (g g \rightarrow H),  {\sigma} (gg\to {\tilde 
t 
\tilde{\bar{t}}}\ )$ are the hard  production cross-sections which are given 
later on. The model normalisation constants  $C_{h}, C_{\tilde t \tilde 
{\bar{t}}}$ are fixed from the fit to dijet diffractive production,
and are given 
in Ref.\cite{bialas1, bialas2, dpemc}.

In the model,  the soft pomeron 
trajectory is
taken 
from the standard 
Donnachie-Landshoff   parametrisation \cite{pom},
 namely $\alpha(t) = 1 + 
\epsilon + \alpha't$, with
$\epsilon \approx 0.08$ and $\alpha' \approx 0.25 
\mathrm{GeV^{-2}}$. 
$\lambda_{h}, \lambda_{\tilde t \tilde{\bar{t}}}$ are  
kept as in  the original paper \cite{bialas1,bialas2} for the SM Higgs boson 
and 
$q 
\bar q$ pairs \footnote{The expression of $C_{h}, C_{t  
{\bar{t}}}$ are explicitely given in formulae (2.13) and (2.14) of
Ref.\cite{bialas1} and formulae (4) and (15) from Ref.
\cite{bialas2}, and included in DPEMC \cite{dpemc}.
}.  Note again that, in this model, the 
strong (non perturbative) 
coupling constant is fixed to a reference value 
$G^2/4\pi=1.$

In order to select exclusive diffractive states,  one needs to take into 
account the 
corrections from soft hadronic scattering. Indeed, the soft 
scattering  
between incident particles tends to mask the genuine
hard diffractive interactions at 
hadronic colliders. The formulation of this 
correction \cite{sp,Kupco:2004fw}  leads to
an overall  correction factor of 3\% which is.
the commonly used correction factor  \cite{khoze,us} for the QCD 
exclusive diffractive  processes at the LHC.

\subsection{Experimental context}
The {\tt DPEMC} \cite{dpemc} Monte Carlo program provides 
an implementation of the Higgs boson, top squark and bottom squark pair
production described above in
both exclusive and inclusive double pomeron exchange modes.
It uses {\tt HERWIG} \cite{herwig} as a cross-section library of
hard QCD 
processes and, when required, convolutes them with the relevant pomeron 
fluxes and parton densities. The survival probabilities discussed in the 
previous section 
(0.03 for double pomeron exchange 
processes)
have been introduced at the generator level. The cross-sections at the 
generator
level are given in the next section after this effect is taken into account.

A possible experimental setup for forward proton detection is described in 
detail in
\cite{helsinki, us}. We will only describe its main features here and discuss 
its
relevance for the Higgs boson and top squark or bottom squark mass 
measurements.

In exclusive DPE or QED processes at the LHC, 
the mass of the central heavy object can be reconstructed
using the roman pot detectors and tagging both protons
in the final state. It is given  by $M^2 = \xi_1\xi_2 s$, where 
$\xi_i$ are 
the proton fractional momentum losses, and $s$  is the total centre-of-mass 
energy squared. 

In the following, we  assume the existence of two detector stations, located at 
 approximately
$210$~m and $420$~m \cite{helsinki} from the interaction point. 
The $\xi$ acceptance and 
resolution have been derived for each device using a complete simulation
of the LHC beam parameters~\cite{helsinki}. The combined $\xi$ acceptance is close to $\sim 
60\%
$ at low masses of about 100~GeV, and 90\% at higher masses
starting at about 220~GeV   
for $\xi$ ranging from $0.002$ to $0.1$. In particular, this means that the
low mass objects (Higgs bosons or top squarks) are mainly detected in the 420 m 
pots
whereas the heavier ones in the closer pots at 210 m
\footnote{Information from the 220~m pots can be included into the
  first level of the trigger. However, the information from the
  pots located at 420~m come too late to reach the first level
  trigger, and can only be used in the second level
 trigger. Therefore, one must use 
  either asymmetric events (one tag at 220~m and another one at
  420~m, where the trigger is based on the information from the 220~m pot) or directly
  a trigger from the main ATLAS and CMS detector. The latter
  option is challenging and under study in both collaborations.}.

Our analysis does not assume any particular value for the $\xi$ resolution. 
We will discuss in Sections III~B and IV~C how the resolution on the Higgs 
boson 
or
the top squark quark masses depend on the detector resolutions, or in other 
words,
the missing mass resolution.

\section{SUSY Higgs boson production}

\subsection{Theoretical aspects}

Let us briefly recall the properties of the lightest MSSM Higgs 
boson $h$
(for recent reviews, see Refs.~\cite{djouadi05} and~\cite{carena02}).
As is well-known, $h$ is constrained to be lighter than the $Z$ boson at 
tree-level.
Once radiative corrections are taken into 
account~\cite{okada91,ellis91,haber91},
the upper limit on its mass
becomes $m_h \lesssim 135$~GeV. The actual value of $m_h$ depends on
several MSSM parameters: two parameters that are sufficient to describe the 
Higgs boson sector
at tree-level, generally chosen to be $m_A$ and $\tan \beta$, the ratio of the 
vacuum expectation values
of the two Higgs boson doublets of the MSSM; and additional parameters that 
control 
the size
of the radiative corrections. These are the top squark and bottom squark soft 
supersymmetry 
breaking
masses, assumed in this paper to be degenerate  and denoted by $M_{SUSY}$,
the top squark and bottom squark
triscalar couplings ($A$-terms) $A_t$ and $A_b$, and the supersymmetric Higgs 
boson 
mass
parameter $\mu$. The dependence of the lightest Higgs boson mass on these 
parameters
can be roughly described as follows: $m_h$ increases with $m_A$ and $\tan 
\beta$,
as well as with the common third generation squark mass $M_{SUSY}$. Its value
also depends strongly on the top squark mixing parameter $X_t \equiv A_t - \mu 
\cot 
\beta$:
starting from the ``minimal mixing'' $X_t = 0$, it increases with $X_t$ and 
reaches
a maximum for $X_t \approx \sqrt{6} M_S$, where
$M^2_S \equiv (m^2_{\tilde t_1} + m^2_{\tilde t_2}) / 2$ is the average of the 
two top squark squared masses ($M_S \simeq M_{SUSY}$ in the limit $M_{SUSY} \gg 
m_t$). This is
illustrated by the following approximate formula for the one-loop upper bound 
on 
the lightest
Higgs boson mass, valid in the decoupling limit $m_A \gg m_Z$ and for
$m_t X_t \ll M^2_S$~\cite{haber96,carena02}:
\be
  m^2_h\ \leq\ m^2_Z \cos^2 2 \beta + \frac{3 g^2 m^4_t}{8 \pi^2 m^2_W}
  \left[ \ln \left( \frac{M^2_S}{m^2_t} \right)
  + \frac{X^2_t}{M^2_S} \left(1 - \frac{X^2_t}{12 M^2_S} \right) \right] .
\ee
In the minimal mixing case, $m_h$ can reach an upper limit of about $120$~GeV 
for
$M_{SUSY} \lesssim 1$~TeV, while it can reach about $135$~GeV in the maximal
mixing case~\cite{carena02}.

The couplings of $h$ can significantly depart from those of the SM Higgs boson.
In particular, its tree-level couplings to down-type and up-type fermions
(normalised to the SM Higgs boson couplings) are given by:
\be
  g_{hff}\ =\ \left\{ \begin{array}{lll} \sin (\beta - \alpha) + \cot \beta\, 
\cos (\beta - \alpha)
    & & \mbox{(f = up-type fermion)}  \\
    \sin (\beta - \alpha) - \tan \beta\, \cos (\beta - \alpha)
    & & \mbox{(f = down-type fermion)} \end{array} \right.\ ,
\label{eq:g_hqq}
\ee
where $\alpha$ is the angle that diagonalises the squared 
mass
matrix of the CP-even Higgs boson and defines the physical CP-even states $h$ and $H$. As for the 
couplings
to the gauge bosons, $hZZ$ and $hWW$, they are suppressed by a factor
$\sin (\beta - \alpha)$ relative to their SM values. In the decoupling regime
$m_A \gg m_Z$~\cite{haber93,gunion02}, in which $A$, $H$ and $H^\pm$
are all much heavier than $h$,
$|\cos (\beta - \alpha)| \leq  {\cal O} (m^2_Z / m^2_A) \ll 1$ and therefore 
the 
couplings
of the lightest MSSM Higgs boson $h$ approach those of the SM
Higgs boson (in particular, $g_{hff} \simeq 1$). On the contrary, if
$m_A \sim m_Z$ (or more precisely $m_A < m^{max}_h$, where $m^{max}_h$ is
the maximal value $m_h$ can reach for fixed values of the squark parameters),
$|\cos (\beta - \alpha)| \sim 1$ and therefore $h$ has significantly different 
couplings
from those of the SM Higgs boson. In particular, at large  $\tan
\beta$, in the ``anti-decoupling'' regime, its couplings to 
down-type 
fermions
are strongly enhanced ($|g_{hbb}| \simeq |g_{h\tau\tau}| \simeq \tan \beta \gg 
1$),
while its couplings to up-type fermions and gauge bosons are suppressed
($|g_{htt}| \sim \cot \beta \ll 1$ and $g_{hWW} = g_{hZZ} = \sin (\beta - 
\alpha) \ll 1$,
in units of the SM Higgs boson couplings).
As we shall see below, this enhances the production cross-section of the
lightest Higgs boson via gluon fusion, while the associated production
with gauge bosons, $q \bar q \rightarrow Z h / W h$, is suppressed. 
Also the partial decay width of $h$ into $b \bar b$ ($\tau^+ \tau^-$),
which is proportional to $g^2_{hbb}$ ($g^2_{h\tau\tau}$), is enhanced.
By contrast, the decay $h \rightarrow \gamma \gamma$, which in
the decoupling regime is dominated by the $W$ boson loop,
does not benefit from such an enhancement at large $\tan \beta$ (the
subdominant bottom quark loop is enhanced, but the dominant $W$
boson loop is suppressed). This decay has therefore a suppressed branching ratio
in the anti-decoupling regime. We close this short review of the
anti-decoupling regime by noting that the heavier CP-even Higgs boson $H$,
in contrast to $h$, has SM-like couplings, but is much heavier than $h$ and 
$A$.
Finally, another regime of interest is the
``intense-coupling'' regime \cite{boos02}, which occurs when
$m_A \sim m^{max}_h$ and
$\tan \beta$ is large. In this regime, all three neutral Higgs bosons are very 
close in mass,
$m_h \approx m_A \approx m_H$, and have enhanced (suppressed) couplings to
down-type fermions (down-type fermions and gauge bosons -- the couplings $AWW$
and $AZZ$ are forbidden by CP invariance), so that it may be difficult to 
distinguish
among them at the LHC. 

Let us now discuss the  production of the lightest MSSM Higgs boson via gluon
fusion~\cite{gunion86,djouadi91,dawson91,spira95,dawson96,harlander03,
harlander03_bis,harlander04} (see Appendix~\ref{sec:appendix} for the relevant
formulae). In the SM, top quark loops give the main contribution
to the cross-section, and bottom loops give a smaller contribution.
In the MSSM, the contribution of the bottom loops can become very large
at large $\tan \beta$ (in the regime where the $hbb$ couplings,
Eq.~(\ref{eq:g_hbb}), are enhanced)
while the top quark loops are suppressed, resulting in an enhancement of the 
gluon fusion cross-section. 
In addition, top and bottom squark loops contribute.
However, top squark loops significantly affect the cross-section only in the 
case of a light top squark, $m_{\tilde t_1} \lesssim (200-400)$~GeV.
In the decoupling regime, their effects are particularly spectacular
in the presence of a large top squark mixing, in which case they interfere
destructively with the top quark contribution~\cite{djouadi98}.
For bottom squark loops to be sizable, a large value of $\tan \beta$ is also 
needed, as well as a large value of $|\mu|$ in the decoupling regime.

In the regime we are interested in, which is characterised by a large value of 
$\tan \beta$ and a suppressed value of $\sin (\beta - \alpha)$, there is no
enhancement of the $h \tilde t \tilde t$ couplings at large top squark mixing,
contrarily to the situation in the decoupling regime.
However, the $h \tilde b \tilde b$ couplings  are enhanced at large
mixing; but this is compensated by the fact that the bottom squark masses
are generally larger than the lightest stop mass $m_{\tilde t_1}$
in typical MSSM scenarios.
Therefore, we neglect the squark loops in the following discussion,
although they are included in our numerical results. Neglecting as well the 
terms
suppressed by $\cot \beta$, we then find the following enhancement factor
for the MSSM cross-section with respect to the SM cross-section
(the QCD corrections to the leading order cross-sections are expected to reduce
this ratio~\cite{djouadi05} by some $30 \%
$ at large $\tan \beta$):
\be
  \frac{\sigma_{MSSM} (g g \rightarrow h)}{\sigma_{SM} (g g \rightarrow h)}\ 
\approx\
    \left| \sin (\beta - \alpha) - \tan \beta \cos (\beta - \alpha)
    \frac{A^h_b (\tau_b)}{A^h_t (\tau_t) + A^h_b (\tau_b)} \right|^2\ ,
\ee
where the loop functions $A^h_t (\tau)$ and $A^h_b (\tau)$ are defined
in Eq.~(\ref{eq:A_h}), and $\tau_{t (b)} \equiv m^2_h / 4 m^2_{t (b)}$.
We therefore expect a large enhancement factor at large values of $\tan \beta$ 
in the
regime where $m_A < m^{max}_h$. This can indeed be seen in Fig. \ref{higgs1}
(upper plot), where
 $\sigma_{MSSM} (g g \rightarrow h) / \sigma_{SM} (g g \rightarrow h)$
 is shown as a function of $m_h$ for $\tan \beta = 30$ and various values of the
 squark parameters. In the maximal mixing case, $m^{max}_h$ is large
 and the antidecoupling condition $m_A < m^{max}_h$  is satisfied over the 
range
 $90\, \mbox{GeV} \leq m_h \leq 120\, \mbox{GeV}$ (remember that in this regime
 $m_h \approx m_A$),  hence $\cos^2 (\beta - \alpha)$ remains very close to 
$1$,
 and the curve essentially reflects the dependence of the loop functions
 $A^h_t (\tau_t)$ and $A^h_b (\tau_b)$ on $m_h$. In the minimal mixing case,
 $m^{max}_h$ is smaller, especially for $M_{SUSY} = 500$~GeV (namely
 $m^{max}_h \approx 114$~GeV for $M_{SUSY} = 1$~TeV, and
 $m^{max}_h \approx 107$~GeV for $M_{SUSY} = 500$~GeV), so that one leaves
 the anti-decoupling regime for much lower values of $m_h$ than in the maximal
 mixing case. This explains why
 the enhancement factor strongly decreases when $m_h$ approaches $m^{max}_h$,
 and finally reaches the decoupling regime value $\sigma_{MSSM} (g g 
\rightarrow 
h) / \sigma_{SM} (g g \rightarrow h) = 1$ for $m_h = m^{max}_h$ (up to squark 
loop effects,
 which remain small for the squark parameters considered here).
Fig. \ref{higgs1} (lower plot) shows the dependence of the enhancement factor 
on 
$\tan \beta$
for $m_h = 100$~GeV. For this value of $m_h$, the condition $m_A < m^{max}_h$
is satisfied for all four sets of the squark parameters considered. For 
moderate 
values
of $\tan \beta$ ($\tan \beta < 5$ is excluded experimentally for $m_h = 100$ 
GeV),
the anti-decoupling regime is not yet reached, \emph{i.e.}, $\cos^2 (\beta 
-\alpha)$
is large but not maximal ($\cos^2 (\beta -\alpha) \sim 1$).
For larger values of $\tan \beta$, e.g., $\tan \beta \gtrsim 20$,
the anti-decoupling regime is reached and the enhancement factor grows with
$\tan^2 \beta$, as expected.

\subsection{SUSY Higgs boson production at the LHC}
\begin{figure}
  \includegraphics[width=0.85\linewidth]{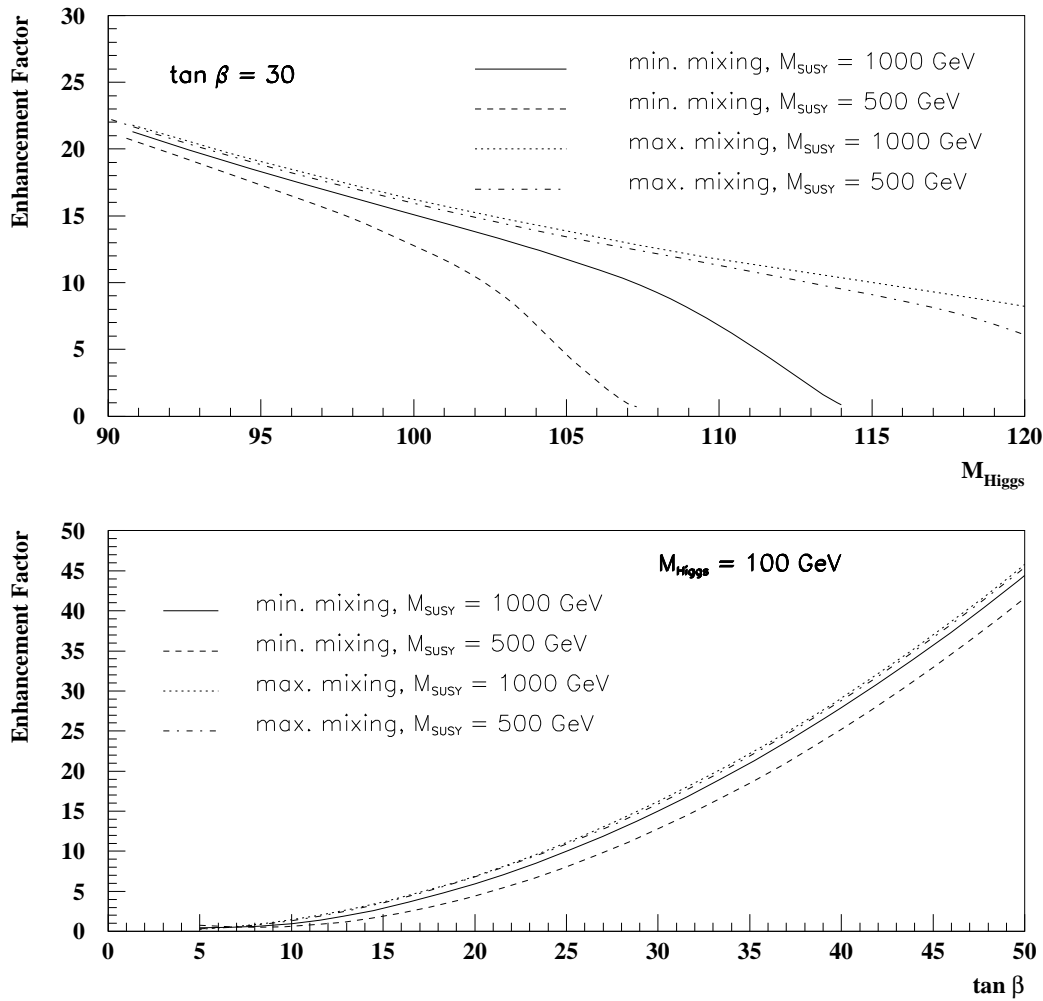}
  \caption{
    Enhancement factor for the diffractive Higgs boson production cross 
section.
    Upper plot: enhancement factor as a function of the Higgs boson mass
    for a value of $\tan \beta=$ 30, and different mixing scenarii and 
    $SUSY$ masses (full line: minimal mixing, $M_{SUSY}=1000$~GeV;
    dashed line: minimal mixing, $M_{SUSY}=500$~GeV;
    dotted line: maximal mixing, $M_{SUSY}=1000$~GeV;
    dashed dotted line: maximal mixing, $M_{SUSY}=500$~GeV).
    Lower plot: similar study as a function of $\tan \beta$ for a Higgs boson
    mass of 100~GeV. The MSSM Higgs boson spectrum has been obtained
    using the program FeynHiggs~\cite{FeynHiggs}, with $\mu = 200$ GeV,
    $M_3 = 800$ GeV, $M_2 = 200$ GeV and $M_1 = (\alpha_1 / \alpha_2) M_2$.
    }
  \label{higgs1}
\end{figure}

In this section, we address the MSSM Higgs boson production for masses
below 120~GeV, when the Higgs boson decays into $b \bar{b}$, the least 
favourable
case at the LHC.
As mentioned in the previous section,  Figure~\ref{higgs1} shows the
cross 
section enhancement factor for SUSY
Higgs boson production with respect to the Standard Model case at generator
level. In the upper plot of Fig. \ref{higgs1}, 
%% we show the dependence of the
%% enhancement factor as a function  of the Higgs boson mass for four different
%% scenarii for $\tan \beta=$ 30. 
the full and dashed lines show the results for the minimal mixing
scenario for two common values of the mass, $M_{SUSY}$,  of third generation MSSM squarks 
(1000 and 500~GeV).
They lead to typical masses of the top squark and bottom squarks of 1010 or
520~GeV, respectively. The cross-section was computed using bottom,
top, top squark and bottom squark  loops, while the effect of top squark 
and 
bottom squark loops is
less than one per mil. The enhancement factor can go up to
a factor 20 compared to the Standard Model case, but is very dependent on the
mass of the Higgs boson. In the maximal smearing scenario (dotted and dashed 
dotted curves in Fig. \ref{higgs1}), the enhancement factor is found to be
similar to that at low Higgs boson masses, but remains important at higher 
masses.

The bottom plot of Fig. \ref{higgs1} displays  the dependence as a function of 
$\tan 
\beta$
for a Higgs boson mass of 100~GeV and  for the same scenarios as before.
The enhancement factor for the Higgs boson production cross
section can reach a factor of up to 45 for a value of $\tan \beta$ of 50.
For this particular value of the mass of the Higgs boson, the model dependence
is not very large.

It is important to note that searches will benefit directly from the increase of the
cross-section  since they will be looking for Higgs bosons decaying into 
$b \bar{b}$ in the main detector, and the branching ratio 
of $h \rightarrow b \bar{b}$ is quite stable
as a function of the MSSM parameters in this region of phase space \footnote{
This is not the case when one looks into non diffractive MSSM Higgs bosons 
decaying
into $\gamma \gamma$ which is strongly suppressed at high $\tan \beta$, see
paragraph III A.
}. Thus, the 
search for diffractive production of MSSM Higgs bosons is the only one benefiting fully 
from the
increase of the cross-section at high values of $\tan \beta$.

\begin{figure}
  \includegraphics[width=0.55\linewidth]{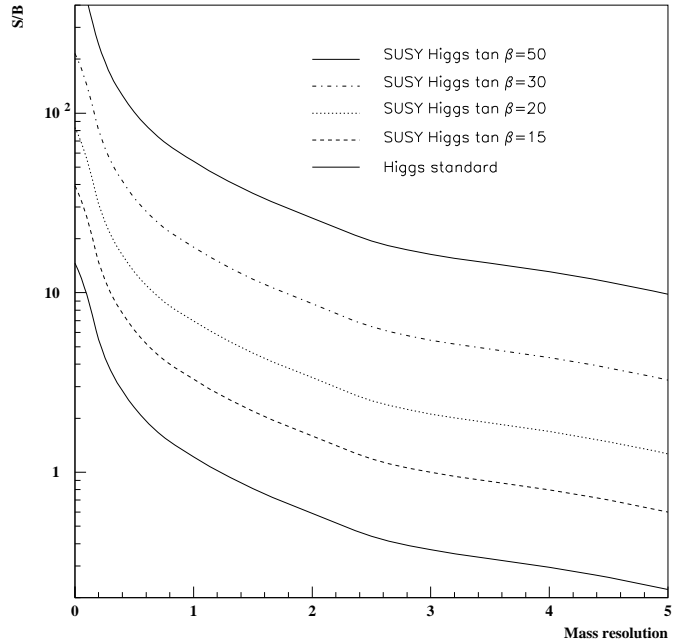}
  \caption{Signal over background as a function of roman pot mass 
  resolution for a Higgs boson mass of 120~GeV and  for different values of 
  $\tan \beta$. From bottom to top: full line: Standard Model Higgs boson,
  dashed line: SUSY Higgs boson with $\tan \beta =$15, dotted line: $\tan \beta 
=$20, 
  dashed dotted line: $\tan \beta =$30,  full line: $\tan \beta =$50.}
  \label{higgs2}
\end{figure}

In the following, we perform a detailed study of
signal-over-background ratio in the case of a Higgs boson mass of
120~GeV. We chose this particular mass since in most of the MSSM
parameter space, the Higgs boson mass is below this value, and this
mass leads to the least favourable scenario (the lowest cross-section
and signal-over-background ratios) with respect to lower masses.  In
Fig. \ref{higgs2}, we give the signal-over-background ratio for
Standard Model and MSSM Higgs boson production for a mass of the Higgs
boson of 120~GeV and for different values of $\tan \beta$, as a
function of the roman pot mass resolution (corresponding to the Higgs boson
mass resolution) for a luminosity of 100 fb$^{-1}$. This study was
performed after a fast simulation of the CMS detector (the ATLAS
detector simulation is expected to produce very similar results) and
experimental cuts described in the following paragraph.

First of all, we require both final state protons  to be
detected in the roman pot detectors, and we take into account the
acceptance of these detectors as it is discussed in section I.  The
cuts applied in the analysis are detailed in Ref. \cite{us1}. The basic
idea is to require two high $p_T$ b-jets with $p_{T1} >$ 45~GeV,
$p_{T2} >$ 30~GeV, originating from the decay of the Higgs boson into $b
\bar{b}$ at low masses. The difference in azimuth between the two jets
should be $170 < \Delta \Phi < 190$ degrees, asking the jets to be
back-to-back.  Both jets are required to be central, $|\eta| <$ 2.5,
b-tagged, with the difference in rapidity of both jets satisfying
$|\Delta \eta| < 0.8$. A cut is applied on the ratio of the dijet
mass to the total mass of all jets measured in the calorimeters,
$M_{JJ}/M_{all} >$ 0.75. The ratio of the dijet mass to the missing
mass should fulfil $M_{JJ} / (\xi_1 \xi_2 s)^{1/2} >$~0.8.

The case for the Standard Model Higgs boson was already given in
\cite{us1}, and we follow the same approach concerning the background
and signal studies.  To compute the signal over background ratios,
both signals and backgrounds dominated by exclusive $b \bar{b}$
production have been integrated over a 2~GeV mass window. After cuts,
the typical number of events expected for the signal of a 120~GeV
Higgs boson and for a luminosity of 100 fb$^{-1}$ is 27.1, 73.2, 154,
398 and 1198 for Standard Model and MSSM ($\tan \beta$= 15, 20, 30 and
50) Higgs boson production.  If the Higgs boson is supersymmetric and
if $\tan \beta$ is large, the diffractive production of MSSM Higgs
bosons could lead to a discovery in the double pomeron exchange mode
at the LHC.  Figure~\ref{higgs2} demonstrates that the signal over
background can reach a value up to 54, 26, 16, and 13 for respective
Higgs boson mass resolutions in roman pot detectors of 1, 2, 3 and
4~GeV and for a value of $\tan \beta$ of 50 for a luminosity of 100
fb$^{-1}$.

\section{Production of top  (bottom) squark pairs}

\subsection{Theoretical framework}

In the MSSM, for each quark flavour $q$, there are two supersymmetric scalar 
partners
$\tilde q_L$ and $\tilde q_R$ associated with the two fermion chiralities $q_L$ 
and $q_R$.
In general, these scalars are not mass eigenstates, due to the presence of soft 
supersymmetry
breaking terms which mix them, the $A$-terms $A_q y_q \tilde Q_L \tilde 
q^\star_R 
H
+ \mbox{h.c.}$, where $y_q$ is the Yukawa coupling of the quark $q$ and $H$ is 
one of the
two MSSM Higgs boson doublet. The mass eigenstates $\tilde q_1$ and $\tilde 
q_2$ 
are
obtained by diagonalising the following $2 \times 2$ matrix \cite{ellis83}:
\be
 \left( \begin{array}{cc} M^2_Q + m^2_q + D_L  &  m_q X_q  \\
                                         m_q X_q  &  M^2_R + m^2_q + D_R
                                         \end{array} \right) ,
\ee
where $D_L \equiv (T^3_q - Q_q  \sin^2 \theta_W ) m^2_Z \cos 2 \beta$,
$D_R \equiv Q_q  \sin^2 \theta_W m^2_Z \cos 2 \beta$, $X_q \equiv A_q - \mu 
\cot 
\beta$
for up-type squarks and $X_q \equiv A_q - \mu \tan \beta$ for down-type 
squarks.
The soft-supersymmetry-breaking squark masses $M_Q$ and $M_R$ are
of the order of the supersymmetry-breaking scale $M_{SUSY}$, and 
phenomenological
constraints require the $A$-term parameter $A_q$ to be at most a few
times $M_{SUSY}$
\cite{CCB}.
Neglecting the terms proportional to $m^2_Z$ in the squark mass matrix, the 
mass
eigenvalues and the mixing angle, which relates the weak interaction eigenstates
$\tilde q_{L,R}$ to the mass eigenstates $\tilde q_{1,2}$, are given by the 
following
expressions:
\be
  m^2_{\tilde q_{1,2}}\ =\ \frac{1}{2} \left( M^2_Q + M^2_R + m^2_q
  \mp \sqrt{(M^2_Q - M^2_R)^2 + 4 m^2_q X^2_q}\, \right) ,  \qquad
  \tan \theta_{\tilde q}\ =\ \frac{2 m_q X_q}{M^2_Q - M^2_R}\ .
\ee
In practise, the mixing is significant only if $m_q X_q \sim M^2_{SUSY}$, which 
can
occur for the top squark and, at large $\tan \beta$, also for the bottom squark. In 
this case,
one can have a strong hierarchy between the two mass eigenstates,
$m_{\tilde q_1} \ll m_{\tilde q_2}$. The $95 \%
$ C.L. experimental bounds on the 
lightest top squark
and bottom squark masses are $m_{\tilde t_1} > 95.7$~GeV and $m_{\tilde b_1} > 
89$ 
GeV,
respectively, while the bound on the other squarks is $250$~GeV \cite{PDG04} 
(the latter
bound also applies to $\tilde b_1$ if mixing effects are small in the bottom 
squark 
sector).
Although these bounds where derived under specific assumptions and may
therefore not hold in some regions of the MSSM parameter space, they are
rather robust.

In our experimental study, we consider the following values for the 
lightest top squark mass:
$174.3$~GeV (\emph{i.e.}, $m_{\tilde 
t_1} = m_t$),
$210$~GeV and $393$~GeV. As we will see, the resolution that can be obtained
on the top squark mass crucially depends on its decay width, which in turn is a 
function of
the top squark mass and of the other MSSM parameters. If the top squark is very 
light,
it is likely to be the next-to-lightest supersymmetric particle, assuming the 
lightest
supersymmetric particle (LSP) is the lightest neutralino $\tilde \chi^0_1$.
Then all two-body decay channels occurring at the tree level are closed. If in 
addition
the tree-body decay channels $\tilde t_1 \rightarrow b W^+ \tilde \chi^0_1$ and
$\tilde t_1 \rightarrow b H^+ \tilde \chi^0_1$ are kinematically not
accessible,
the main decay mode is expected to be the loop-induced flavour violating decay
$\tilde t_1 \rightarrow c \tilde \chi^0_1$  \cite{hikasa87}. The decay width of 
the lightest top squark
is then given by \cite{hikasa87}:
\be
  \Gamma (\tilde t_1 \rightarrow c \tilde \chi^0_1)\ =\ \frac{g^2}{16 \pi} 
|f^c_{L1}|^2 |\epsilon|^2
  m_{\tilde t_1} \left( 1 - \frac{m^2_{\tilde \chi^0_1}}{m^2_{\tilde t_1}} 
\right)^2\ ,
\ee
where $|f^c_{L1}| \leq 1$ is the $\tilde c_L$-$c_L$-$\tilde \chi^0_1$ coupling, 
and $\epsilon$
is a flavour-violating insertion. The authors of Ref. \cite{hikasa87} estimated
$|\epsilon| \sim (1-4) \times 10^{-4}$ in mSUGRA, yielding
$\Gamma (\tilde t_1 \rightarrow c \tilde \chi^0_1) \lesssim (0.085-1.4) \times 
10^{-9}\,
m_{\tilde t_1} [ 1 - (m_{\tilde \chi^+_1} / m_{\tilde t_1})^2 ]^2$;
but depending on the mSUGRA parameters $|\epsilon|$ could be either much 
smaller
or larger, in particular at large $\tan \beta$ where $|\epsilon|$ behaves like 
$\tan^2 \beta$.
In non-minimal SUGRA models, $|\epsilon|$
could even be of order one. However, in the regions of the MSSM parameter space 
where
$\Gamma (\tilde t_1 \rightarrow c \tilde \chi^0_1)$ is suppressed, the 
four-body
decay modes $\tilde t_1 \rightarrow b \tilde \chi^0_1 f \bar f$ are likely to 
be
dominant \cite{boehm00,djouadi01}.
For larger top squark masses, the three-body decay channels
$\tilde t_1 \rightarrow b W^+ \tilde \chi^0_1$ and $\tilde t_1 \rightarrow b 
H^+ 
\tilde \chi^0_1$
\cite{porod97,porod99} (and, if the sleptons are lighter than the lightest top 
squark,
$\tilde t_1 \rightarrow b\, \nu_l\, \tilde l^+$ and $\tilde t_1 \rightarrow b\, 
\tilde \nu_l\, l^+$
\cite{hikasa87,porod99,datta99}) open up and tend to dominate.
Finally, when $m_{\tilde t_1} > m_b + m_{\tilde \chi^+_1}$ and
$m_{\tilde t_1} > m_t + m_{\tilde \chi^0_1}$, the two-body decays
$\tilde t_1 \rightarrow b \tilde \chi^+_1$ and $\tilde t_1 \rightarrow t \tilde 
\chi^0_1$
become kinematically accessible and dominate the lightest top squark 
decays\footnote{We
do not consider the decays $\tilde t_1 \rightarrow t \tilde g$ and
$\tilde t_1 \rightarrow \tilde b_i W^+$, $\tilde b_i H^+$, since the gluinos 
and
the other squarks are generally heavier than the lightest top squark.} 
\cite{2_body}.
The partial decay width of $\tilde t_1 \rightarrow b \tilde \chi^+_1$ is given 
by:
\be
  \Gamma (\tilde t_1 \rightarrow b \tilde \chi^+_1)\ =\ \frac{g^2}{16 \pi}\, 
(l^2_{11} + k^2_{11})\,
  m_{\tilde t_1} \left( 1 - \frac{m^2_{\tilde \chi^+_1}}{m^2_{\tilde t_1}} 
\right)^2\ ,
\ee
where $l_{11}$ and $k_{11}$ are chargino couplings. Since $|l_{11}|$,
$|k_{11}| \lesssim 1$, $\Gamma (\tilde t_1 \rightarrow b \tilde \chi^+_1)
\lesssim 0.0085\, m_{\tilde t_1} [ 1 - (m_{\tilde \chi^+_1} / m_{\tilde t_1})^2 
]^2$.
The decay $\tilde t_1 \rightarrow t \tilde \chi^0_1$ has a larger phase space 
suppression.

For our experimental study, we do not consider any specific benchmark scenario, 
but simply assume $m_{\tilde t_1} < m_{\tilde \chi^+_1}$ for the cases 
$m_{\tilde t_1} = 174$ and $210$~GeV. Then, we conservatively take 
$\Gamma_{\tilde 
t_1} = 100$ MeV
for $m_{\tilde t_1} = 174$ and $210$~GeV, although the actual decay width could 
be 
much 
smaller.
For $m_{\tilde t_1} = 393$~GeV, we assume that the two-body decay
$\tilde t_1 \rightarrow b \tilde \chi^+_1$ is accessible, and we take the decay 
width
computed for SPS 1a, $\Gamma_{\tilde t_1} = 1.8$~GeV \cite{ghodbane02}.

\subsection{Stop production cross-section and missing mass distribution}

At hadron colliders, top squarks can be produced at lowest QCD order via 
quark-antiquark
annihilation and gluon-gluon fusion. In the present study, we are interested in 
the $J_z=0$, colour-singlet gluon-gluon fusion cross-section; 
it reads, at the parton level~\cite{durhamtop}:
\ba
%%%%%%  \sigma_{LO} (q \bar q \rightarrow \tilde t_i \bar{\tilde t}_i) & = &
%%%%%%    \frac{\alpha^2_S \pi}{s}\ \frac{2}{27}\ \beta^3_i\ ,  \\
  \frac{d\sigma_{LO}}{dt} (g g \rightarrow \tilde t_i \bar{\tilde t}_i) & = &
        \frac{4\pi}{12}\frac{\alpha_s^2}{s^2}\frac{m_{\tilde t_i}^4}{E_T^4}
\ea
where $\sqrt{s}$ is the centre-of-mass energy of the subprocess, $m_{\tilde 
t_i}$ is 
the top squark mass, and $E_T$ is the transverse energy of 
the final particles.

\begin{figure}
  \includegraphics[width=0.55\linewidth]{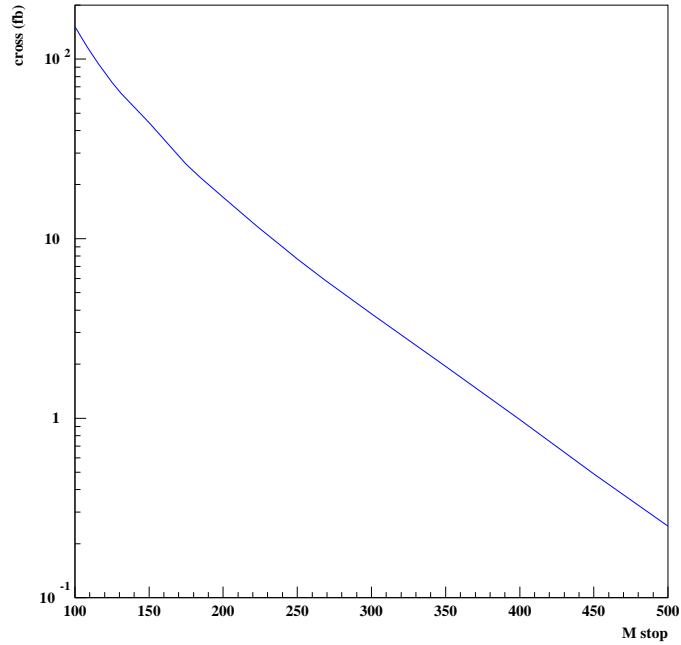}
  \caption{
    Top squark pair production cross-section as a function of the top squark 
mass.}
  \label{stop0}
\end{figure}

The top squark production cross-section has been obtained using the {\tt DPEMC} 
generator \cite{dpemc} after applying a survival probability of 0.03.
The top squark pair production cross-section as a function of the top squark 
mass
is given in Fig. \ref{stop0}. The $\tilde{t} \tilde{\bar{t}}$ production cross 
section
is found to be 26.3, 14.1 and 1.1 fb for a top squark mass of 174.3
(at about the top quark mass), 210 and 393~GeV respectively.

\begin{figure}
  \includegraphics[width=0.55\linewidth]{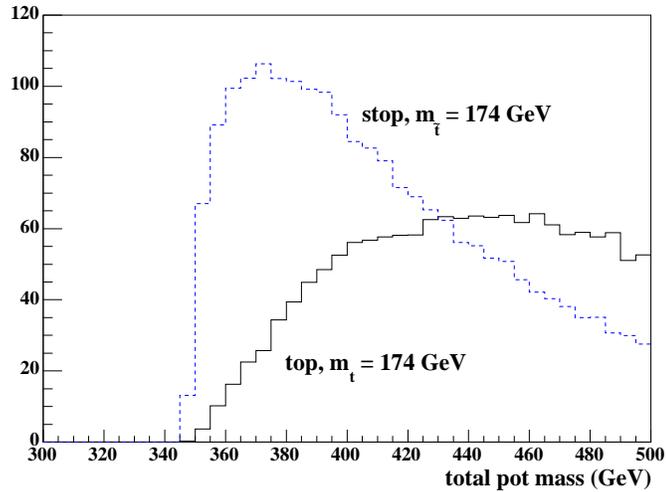}
  \caption{Distribution of the missing mass for 100 fb$^{-1}$ for
  $t \bar{t}$ events (dashed line), and for $\tilde{t} \tilde{\bar{t}}$ 
  (full line) for $m_{top} = m_{\tilde{t}}$. The faster rise of the stop quark 
cross-section  as
  a function of missing mass is 
 due to the scalar nature of  these particles.}
  \label{stop1}
\end{figure}

The distribution of the missing mass distribution for $t \bar{t}$ and 
$\tilde{t} \tilde{\bar{t}}$ events for $m_{\tilde{t}} = m_{t}=$174.3
GeV is shown in Fig. \ref{stop1} for a luminosity of 100
fb$^{-1}$. The missing mass distribution for top squark (top) events is in
full (dashed) line. The cross-section rise at threshold 
is much faster than for top quarks and typical of pair production of scalar 
particles. 
The next section describes a method to determine the stop quark mass
by performing a threshold scan of the missing mass of the $\tilde{t}
\tilde{\bar{t}}$ process, measured with roman pot detectors.

\subsection{Stop mass measurement}
In this section, we describe briefly the method we used to obtain the 
stop mass resolution and its results. The histogram method
\footnote{In Ref. \cite{us2}, we give two methods to measure the $W$ boson or 
the
top mass, namely the histogram or the turn-on fit methods. For a matter of
simplicity, we used only the histogram method in this paper.} is described in 
more 
detail in Ref. \cite{us2}. It compares
the mass distribution in data with some reference distributions following
a Monte Carlo simulation of the detector with different input masses
corresponding to the data luminosity. As an example, we can produce 
a data sample for 100 fb$^{-1}$ with a top squark mass of 210~GeV, and a few 
MC samples corresponding to top squark masses between 180 and 240~GeV by steps 
of
1~GeV. To evaluate the statistical uncertainty due to the method itself,
we perform the fits with some 100 different ``data" ensembles.
For each ensemble, one obtains a different 
reconstructed top squark mass, the dispersion corresponding only to statistical
effects. The $\chi^2$ is defined using the approximation of poissonian errors
as given in Ref.~\cite{Gehrels:1986mj}.  Each ensemble thus gives a $\chi^2$ 
curve
which in the region of the minimum is fitted with a fourth-order
polynomial. The position of the
minimum of the polynomial gives the best value of the top squark mass
and the uncertainty $\sigma(m_{\tilde{t}})$ is obtained from the values where
$\chi^2 = \chi^2_\text{min} + 1$. 

\begin{figure}
  \includegraphics[width=0.55\linewidth]{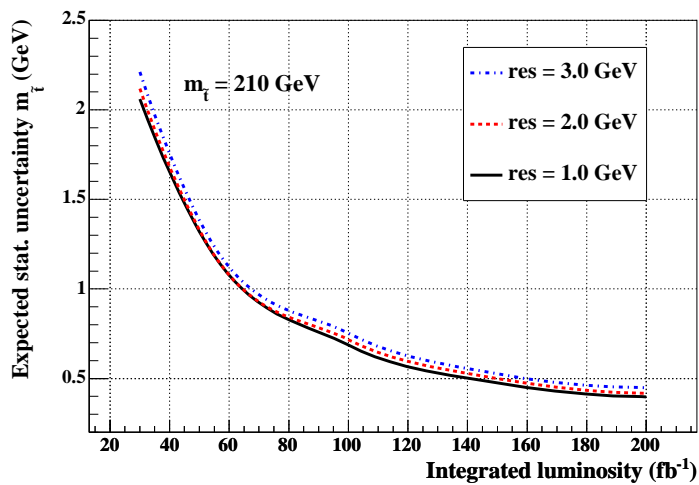}
  \caption{
    Expected statistical precision of the $\tilde{t}$ mass as a function of the
    integrated luminosity for various resolutions of the roman pot
    detectors using the histogram-fitting method.}
  \label{stop2}
\end{figure}

\begin{figure}
  \includegraphics[width=0.55\linewidth]{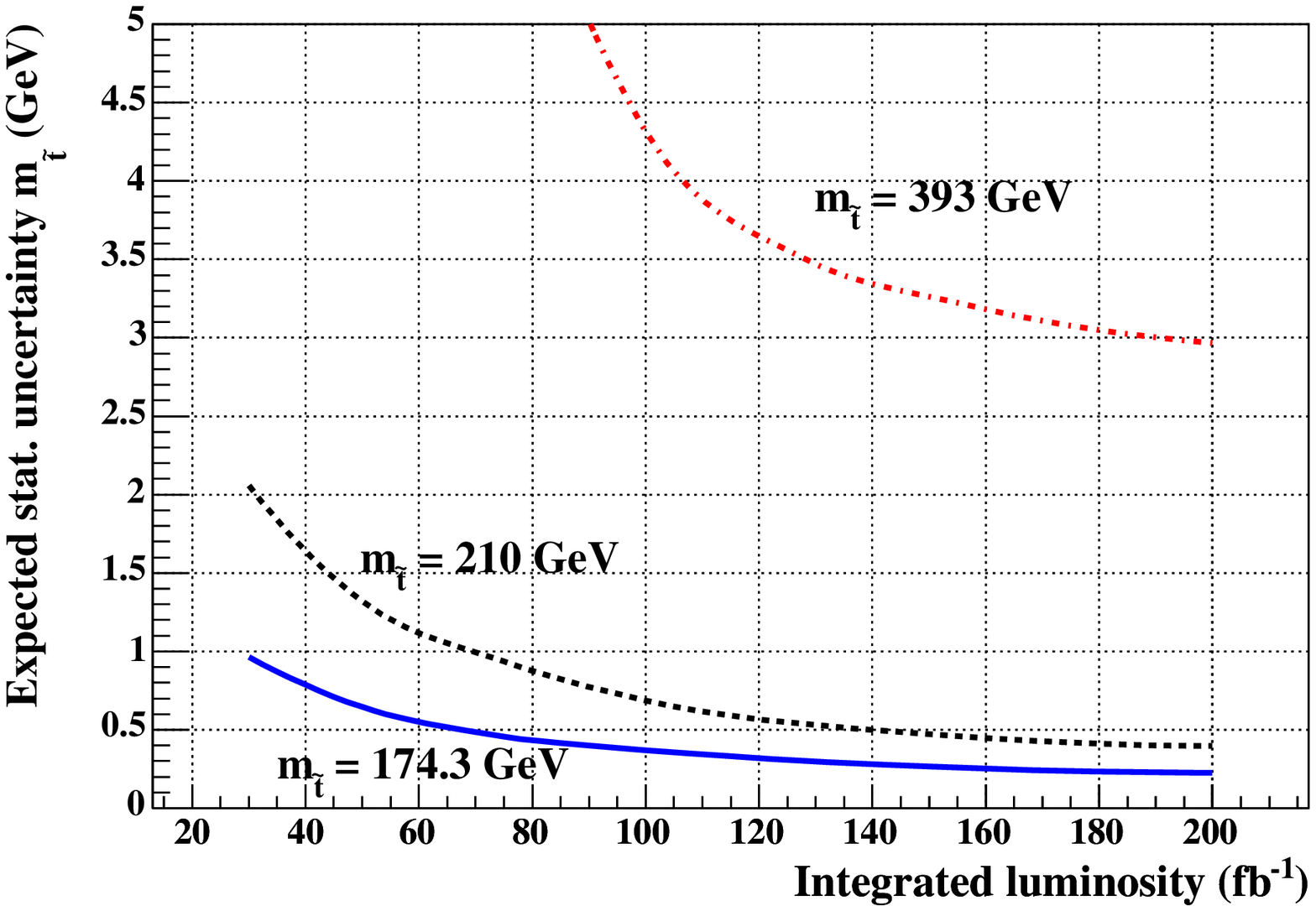}
  \caption{
    Expected statistical precision of the $\tilde{t}$ mass as a function of the
    integrated luminosity for different $\tilde{t}$ masses (174.3, 210 and 
    393~GeV).}
  \label{stop3}
\end{figure}

The results are given in Fig. \ref{stop2} and \ref{stop3}. Fig. \ref{stop2}
displays the results on the top squark mass resolution for a top squark mass of 
210~GeV as 
an
example, as a function of luminosity, for different roman pot resolutions.
The results depend only weakly on the roman pot resolution and
mostly on the number of events produced for a given luminosity. The resolution
on the top squark mass is thus dominated by statistics. We also note 
that 
the
integrated luminosity does not take into account the efficiency of the cuts to
select the $\tilde{t} \tilde{\bar{t}}$ events since these efficiencies depend 
strongly
on the SUSY parameters. A typical efficiency of 60\% is found requesting
a missing transverse energy to be greater than 80 GeV, and either two
reconstructed jets or one lepton and one jet with a transverse momentum
greater than 20 GeV.
Figure~\ref{stop3} displays the resolution obtained for the three values of
the top squark mass discussed above. A resolution of
about 0.4, 0.7 and 4.3~GeV is obtained for a top squark mass of 174.3, 210 and 393~GeV for 
a luminosity (divided by the signal efficiency) of 100 fb$^{-1}$. As it was 
mentionned 
in paragraph IV A, the top squark width has been taken into account in this 
study. 
For a top squark mass of 174.3, 210 GeV, the top squark width of 100 MeV
has a negligible effect, whereas the top squark width of 1.8 GeV for a top 
squark mass
of 393 GeV cannot be neglected.

\section{Conclusion and outlook}
In this paper, we described the advantages of diffractive SUSY particle 
productions 
for two different processes, namely the MSSM Higgs bosons and top squark pairs.
The  cross-section for diffractive MSSM Higgs bosons production is noticeably 
enhanced 
at high values of $\tan \beta$ and since we look for Higgs boson decaying into 
$b
\bar{b}$, it is possible to benefit directly from the enhancement of the cross
section contrary to the non diffractive case. A signal-over-background up to a
factor 50 can be reached for 100 fb$^{-1}$ for $\tan \beta \sim 50$. In 
particular, we analysed in detail the {\it antidecoupling} regime,  in which  
$H$ behaves like the SM Higgs boson, while $h$ has enhanced (resp. suppressed) 
couplings to down-type fermions (resp. up-type fermions and gauge 
bosons). We find that central diffraction production seems to be promising in 
that regime. 

%%%%%%%%%%%%%%%%%%%%%%%%%%%%%%%%%%%%%%%%%%%%%%%%%%%%%%%%%%%%%%%%%%%%%%%%%%%

The other application is to use the so-called ``threshold-scan method"
to measure the top squark mass in {\it exclusive} events. The idea is quite 
simple: 
one
measures the turn-on point in the missing mass distribution above twice
the top squark mass. After taking into account the top squark width, we obtain 
a 
resolution
on the top squark mass of 0.4, 0.7 and 4.3 GeV for a top squark mass of 174.3, 
210 and 393
GeV for a luminosity (divided by the signal efficiency) of 100
fb$^{-1}$ and the production rates calculated in Section IV of this
paper. If these rates hold, the typical mass resolutions are
comparable to those  at a linear
collider. The process is thus similar to  those at linear colliders (all final 
states
are detected), but without the initial state radiation problem. 

It should be stressed once more that  production via the diffractive 
 exclusive processes (\ref{exc}) is model dependent, and definitely needs
the Tevatron data to test the models. It will allow to determine more precisely 
the production cross-section by testing and measuring at the Tevatron the jet 
and photon production for high masses and high dijet or diphoton mass fraction. 
If, for instance, we compare (as much as possible since the methods of 
evaluation 
are not exactly comparable) the expectations of   the  `Pomeron-induced' 
\cite{us1,us} and   `proton-induced'
\cite{khoze} models for exclusive 
production, we find striking differences for high mass states, in particular 
the 
stop quark production. 
Indeed, the Sudakov form factors present in the   
`proton-induced' approach induce negligible stop quark cross-sections 
\cite{durhamtop}. 
The search for exclusive events in high mass dijet or diphoton events at the LHC is of
considerable importance to test the models.
For MSSM Higgs bosons in the low mass range there is not an
order-of-magnitude difference between the two models as was already mentioned 
for the standard Higgs production \cite{forshaw}. 

On the other hand, it is also possible to perform a similar study using
{\it inclusive} double pomeron exchanges (\ref{inc}). These processes have 
already been 
observed by many experiments but suffer from the lack of  knowledge on the 
gluon 
density in the pomeron at high $\beta$. The first step is thus to measure this 
gluon density by, for instance, using dijet events or the threshold scan method 
for  inclusive $t \bar{t}$ production. Once the high-$\beta$ gluon  better
determined, it is possible to look for top squark events,  again using
the 
threshold scan
method and deviation at high masses provided the cross-section is high enough.
This study goes beyond the purpose of the present  paper and it certainly deserves 
a dedicated study \cite{us2}.

\section{Acknowledgements}
We thank Pavel Demine for his  participation in this study at an early stage 
and 
 for useful discussions.

\appendix
\section{MSSM Higgs boson production via gluon fusion at leading order}
\label{sec:appendix}

The cross-section for the lightest MSSM Higgs boson production
via gluon fusion reads, at leading order~\cite{gunion90}:
\be
  \sigma (g g \rightarrow h)\ =\ \frac{G_F \alpha^2_S}{288 \sqrt{2} \pi}
    \left| \frac{3}{4} \sum_q g_{hqq} A^h_q (\tau_q)
    + \frac{3}{4} \sum_{\tilde q} \frac{g_{h \tilde q \tilde q}}{m^2_{\tilde 
q}}
    A^h_{\tilde q} (\tau_{\tilde q}) \right|^2\ ,
\label{eq:gg_to_h}
\ee
where the first term contains the quark loop contributions, and the second
term the squark loop contributions.
%
%$g_{hqq}$ and $g_{h \tilde q \tilde q}$
%denote the coupling of the lightest MSSM Higgs boson $h$ to the quark $q$
%and to the squark $\tilde q$, respectively. 
%
The loop functions $A^h_q (\tau)$ and $A^h_{\tilde q} (\tau)$ are given by:
\be
  A^h_q (\tau)\ =\ \frac{2}{\tau^2} \left[ \tau + (\tau - 1) f (\tau) \right] ,  
\qquad
  A^h_{\tilde q} (\tau)\ =\ \frac{1}{\tau^2} \left[ f (\tau) - \tau \right] ,  
\label{eq:A_h}  \\
\ee
\ba
  f (\tau)\ =\ \left\{ \begin{array}{cll} \arcsin^2 (\sqrt{\tau}) & & \tau \leq 
1  \\
    - \frac{1}{4} \left[ \ln \left( \frac{1 + \sqrt{1 - 1 / \tau}}{1 - \sqrt{1 
- 
1 / \tau}} \right) - i \pi \right]^2
    & & \tau > 1  \end{array} \right.\ ,
\ea
and the parameters $\tau_q$ and $\tau_{\tilde q}$ are defined by
$\tau_q \equiv m^2_h / 4 m^2_q$ and $\tau_{\tilde q} \equiv m^2_h / 4 
m^2_{\tilde q}$, respectively, where $m_q$ (resp. $m_{\tilde q}$) denotes
the mass of the quarks (resp. squarks) running in the loop.
The couplings of the lightest MSSM Higgs boson to the top and bottom quarks,
normalised to the SM Higgs boson couplings, are given by:
\ba
  g_{htt} & = & \sin (\beta - \alpha) + \cot \beta\, \cos (\beta - \alpha)\ ,  
\\
  g_{hbb} & = & \sin (\beta - \alpha) - \tan \beta\, \cos (\beta - \alpha)\ ,
  \label{eq:g_hbb}
\ea
and its couplings to the top squark and bottom squark mass eigenstates,
in units of $g / M_W$, by (we omit the off-diagonal couplings
$g_{h \tilde t_1 \tilde t_2}$ and $g_{h \tilde b_1 \tilde b_2}$, which are not
relevant at leading order):
\ba
  g_{h \tilde t_1 \tilde t_1} & = & - \left( \frac{1}{2} \cos^2 \theta_{\tilde 
t}
    - \frac{2}{3} \sin^2 \theta_W \cos 2 \theta_{\tilde t} \right)\! M^2_Z \sin 
(\beta + \alpha)
    + m^2_t\, \frac{\cos \alpha}{\sin \beta} + \frac{1}{2} \sin 2 
\theta_{\tilde 
t}\, m_t
    \left( A_t \frac{\cos \alpha}{\sin \beta} + \mu \frac{\sin \alpha}{\sin 
\beta} \right) ,\hskip .8cm  \\
  g_{h \tilde t_2 \tilde t_2} & = & - \left( \frac{1}{2} \sin^2 \theta_{\tilde 
t}
    + \frac{2}{3} \sin^2 \theta_W \cos 2 \theta_{\tilde t} \right)\! M^2_Z \sin 
(\beta + \alpha)
    + m^2_t\, \frac{\cos \alpha}{\sin \beta} - \frac{1}{2} \sin 2 
\theta_{\tilde 
t}\, m_t
    \left( A_t \frac{\cos \alpha}{\sin \beta} + \mu \frac{\sin \alpha}{\sin 
\beta} \right) ,  \\ 
  g_{h \tilde b_1 \tilde b_1} & = & \left( \frac{1}{2} \cos^2 \theta_{\tilde b}
    - \frac{1}{3} \sin^2 \theta_W \cos 2 \theta_{\tilde b} \right)\! M^2_Z \sin 
(\beta + \alpha)
    - m^2_b\, \frac{\sin \alpha}{\cos \beta} + \frac{1}{2} \sin 2 
\theta_{\tilde 
b}\, m_b
    \left( - A_b \frac{\sin \alpha}{\cos \beta} + \mu \frac{\cos \alpha}{\cos 
\beta} \right) ,  \\ 
  g_{h \tilde b_2 \tilde b_2} & = & \left( \frac{1}{2} \sin^2 \theta_{\tilde b}
    + \frac{1}{3} \sin^2 \theta_W \cos 2 \theta_{\tilde b} \right)\! M^2_Z \sin 
(\beta + \alpha)
    - m^2_b\, \frac{\sin \alpha}{\cos \beta} - \frac{1}{2} \sin 2 
\theta_{\tilde 
b}\, m_b
    \left( - A_b \frac{\sin \alpha}{\cos \beta} + \mu \frac{\cos \alpha}{\cos 
\beta} \right) ,
\ea
where $\theta_{\tilde t}$ and $\theta_{\tilde b}$ are the mixing angle in the 
stop
and the bottom squark sector respectively, defined by
$\tilde q_1 = \cos \theta_{\tilde q}\, \tilde q_L + \sin \theta_{\tilde q}\, 
\tilde q_R$,
$\tilde q_2 = - \sin \theta_{\tilde q}\, \tilde q_L + \cos \theta_{\tilde q}\, 
\tilde q_R$.

The leading order cross-section for the SM Higgs boson production
via gluon fusion can be obtained from Eq.~(\ref{eq:gg_to_h})
by removing the squark contribution and by setting $g_{hqq} = 1$.

\end{document}